# BIAS COMPENSATION IN ITERATIVE SOFT-FEEDBACK ALGORITHMS WITH APPLICATION TO (DISCRETE) COMPRESSED SENSING

*Susanne Sparrer, Robert F.H. Fischer*

Institute of Communications Engineering, Ulm University, Germany

## ABSTRACT

In all applications in digital communications, it is crucial for an estimator to be unbiased. Although so-called soft feedback is widely employed in many different fields of engineering, typically the biased estimate is used. In this paper, we contrast the fundamental unbiasing principles, which can be directly applied whenever soft feedback is required. To this end, the problem is treated from a signal-based perspective, as well as from the approach of estimating the signal based on an estimate of the noise. Numerical results show that when employed in iterative reconstruction algorithms for Compressed Sensing, a gain of $1.2\,\mathrm{dB}$ due to proper unbiasing is possible.

## 1. INTRODUCTION

In communications engineering, estimation is very often based on the minimum mean-squared error (MMSE) criterion, which is also justified by the relation between MMSE and mutual information [1]. Independent of the problem to be solved and of the application at hand, the used estimators need to be *unbiased*, i.e., no systematic offset has to be present. While the unbiasing is state of the art for some estimators such as linear MMSE estimation [2, 3], for the so-called soft feedback [4], which is used in many different fields such as successive interference cancellation (SIC), a.k.a. decision-feedback equalization (DFE) in multiuser detection [5, 6, 7], unbiasing is generally ignored. In the following, we derive a general rule for the unbiasing of soft feedback, i.e., nonlinear MMSE estimates, and apply it to iterative algorithms in Compressed Sensing.

## 2. UNBIASING OF NONLINEAR MMSE ESTIMATORS

In many problems, an observation[1]

$$z = x + n \quad (1)$$

is available, which can be assumed to be a noisy variant of the true value $x$, with the measurement noise $n$ (variance $\sigma_n^2$) which is independent of $x$. For brevity, all variables are assumed to be zero-mean; a generalization is straightforward. The goal is to estimate $x$ given the knowledge of the distributions of $x$ and $n$, such that the mean-squared error is minimized. To this end, the conditional mean estimator is the optimum solution [8, 9]. In digital communications, the corresponding estimate is often denoted as *soft value* $x_\mathrm{B}$; it is calculated by[2] [8, 4] ($\eta(\cdot)$: real-valued estimator function)

$$x_\mathrm{B} \stackrel{\text{def}}{=} \min_{\tilde\eta} \mathrm{E}\{\|\tilde\eta(z) - X\|_2^2\} = \mathrm{E}_X\{X|z\} \stackrel{\text{def}}{=} \eta_X(z)\,. \quad (2)$$

The estimate can be written as the sum of the true value $x$ and the estimation error $e_\mathrm{B}$, i.e.,

$$x_\mathrm{B} = x + e_\mathrm{B}\,, \quad (3)$$

with variance $\quad \varsigma_{e_\mathrm{B}}^2(z) = \mathrm{var}\{X|z\} = \mathrm{E}_X\{E_\mathrm{B}^2\}\,, \quad (4)$

and mean-squared error $\quad \sigma_{e_\mathrm{B}}^2 \stackrel{\text{def}}{=} \mathrm{E}_Z\{\varsigma_{e_\mathrm{B}}^2(Z)\}\,. \quad (5)$

Note that $\varsigma_{e_\mathrm{B}}^2$ is the squared error averaged over the distribution of $X$ and the noise $N$, thereby keeping the sum of them, i.e., the observation $z$, fixed. $\sigma_{e_\mathrm{B}}^2$, on the other hand, averages $\varsigma_{e_\mathrm{B}}^2$ over all possible values of $Z$.

This estimate, as any MMSE solution, is *biased* (index $\cdot_\mathrm{B}$), i.e., a part of the useful signal is accounted to the error, as for any MMSE solution the error is orthogonal to the estimate $\mathrm{E}_Z\{Z\,E_\mathrm{B}\} = 0$.

### 2.1. Signal-Based Unbiasing

In case of (scalar) linear MMSE estimation, the *biased* estimate is a scaled version of the observation, i.e., $x_\mathrm{B} = k \cdot z$, with the scaling factor $k$. In order for an estimate to be unbiased (index $\cdot_\mathrm{U}$), this scaling has to be compensated for, i.e.,

$$x_\mathrm{U} = h \cdot x_\mathrm{B}\,, \quad (6)$$

where the unbiasing factor $h$ has to be adjusted such that the error $E_\mathrm{U} = X_\mathrm{U} - X$ present in the unbiased estimate is orthogonal to $X$. In the linear case, it follows immediately $h = 1/k$.

---

This work was supported by Deutsche Forschungsgemeinschaft (DFG) under grant FI 982/8-1.

[1]Random variables are denoted by capital letters, actual realizations by lower-case letters.

[2]With an abuse of the term "bit", the soft values are sometimes also denoted as soft bits.

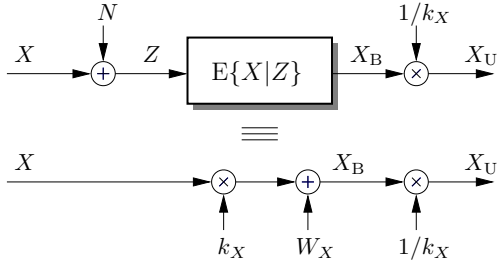

**Fig. 1**. Block diagram of signal-based unbiasing.

While the scaling factor $k$ is obvious for linear estimators, in the case of non-linear estimators an *average* scaling has to be calculated. To this end, the nonlinear estimate $\eta_X(Z)$ is represented as the sum of a linear estimate $k_X X$ and noise $W_X$, which is uncorrelated to the linear estimate [10, 11]

$$X_\mathrm{B} = \eta_X(Z) = k_X X + W_X , \quad (7)$$

as visualized in the block diagram given in Fig. 1, and in Fig. 4. Adjusting $k_X$ according to the MMSE criterion, i.e., such that

$$\mathrm{E}_{XN}\left\{(X_\mathrm{B} - k_X \cdot X)^2\right\} = \mathrm{E}_{XN}\{W_X^2\} \to \min , \quad (8)$$

leads to [11]

$$k_X = \frac{\mathrm{E}_{XN}\{\eta_X(X+N)\,X\}}{\mathrm{E}_X\{X^2\}} = \frac{\mathrm{E}_{XN}\{X_\mathrm{B}\,X\}}{\mathrm{E}_X\{X^2\}} . \quad (9)$$

With $\mathrm{E}_{XN}\{X_\mathrm{B}\,N\} = \sigma_{e_\mathrm{B}}^2$ [14] and $\mathrm{E}_{XN}\{X_\mathrm{B}\,Z\} = \sigma_x^2$, it calculates to

$$k_X = \frac{\mathrm{E}_{XN}\{X_\mathrm{B}\,Z\} - \mathrm{E}_{XN}\{X_\mathrm{B}\,N\}}{\mathrm{E}_X\{X^2\}} = \frac{\sigma_x^2 - \sigma_{e_\mathrm{B}}^2}{\sigma_x^2} = \frac{\sigma_{x_\mathrm{B}}^2}{\sigma_x^2} . \quad (10)$$

Hence, plugging $h = 1/k_X$ into (6) gives

$$x_\mathrm{U} = \frac{\sigma_x^2}{\sigma_x^2 - \sigma_{e_\mathrm{B}}^2} \cdot x_\mathrm{B} = (1 - C_X) \cdot x_\mathrm{B} , \quad (11)$$

where we defined $C_X \stackrel{\text{def}}{=} \sigma_{e_\mathrm{B}}^2 / (\sigma_{e_\mathrm{B}}^2 - \sigma_x^2)$.

The estimate can be written as noisy variant of the true value, i.e., $x_\mathrm{U} = x + e_\mathrm{U}$. As for any unbiased estimate, the (zero-mean) error $e_\mathrm{U}$ is the one with minimum mean-squared error $\mathrm{E}_{XN}\{E_\mathrm{U}^2\}$, which is orthogonal to the signal to be estimated

$$\mathrm{E}_{XE_\mathrm{U}}\{X \cdot E_\mathrm{U}\} = 0 . \quad (12)$$

With straightforward reformulations, the variance of $E_\mathrm{U}$ reads

$$\varsigma_{e_\mathrm{U}}^2 = \mathrm{E}_X\{E_\mathrm{U}^2\} = \mathrm{E}_X\{(x_\mathrm{U} - X)^2\}$$
$$= (1 - C_X^2) \cdot \varsigma_{e_\mathrm{B}}^2 + C_X^2 \cdot \sigma_x^2 . \quad (13)$$

Thus, instead of the variance $\varsigma_{e_\mathrm{B}}^2$ if no unbiasing is applied, a tradeoff between $\varsigma_{e_\mathrm{B}}^2$ and $\sigma_x^2$ is active.

## 2.2. Noise-Based Unbiasing

In the previous section, $X$ has been estimated and unbiased directly. However, since $Z = X + N$, we can also estimate $N$, from which, in turn, $X$ can be calculated. This approach is visualized in Fig. 2 [12].

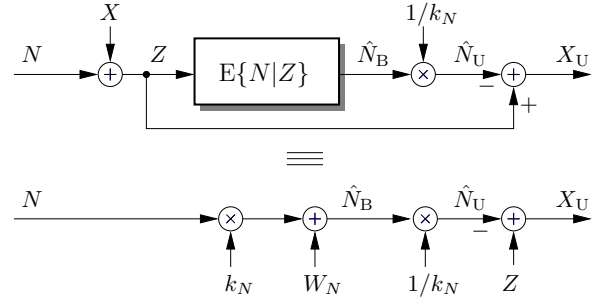

**Fig. 2**. Block diagram of error-based unbiasing.

Note that the nonlinear estimator of the noise is connected to the estimator of the signal by

$$\eta_N(z) \stackrel{\text{def}}{=} \mathrm{E}_N\{N|z\} = \mathrm{E}_N\{(Z-X)|z\} = z - \eta_X(z) . \quad (14)$$

Hence, the output of the estimator can be written as $\hat{N}_\mathrm{B} = Z - X_\mathrm{B}$. Similar to the signal-based case, we linearize the estimator $\eta_N(z)$ by

$$\hat{N}_\mathrm{B} = k_N N + W_N , \quad (15)$$

where, again using $\mathrm{E}_{XN}\{X_\mathrm{B}\,N\} = \sigma_{e_\mathrm{B}}^2$, the scaling factor is calculated by [12, 14]

$$k_N = \frac{\mathrm{E}_{XN}\{\hat{N}_\mathrm{B} N\}}{\mathrm{E}_N\{N^2\}} = \frac{\mathrm{E}_{XN}\{(Z - X_\mathrm{B})N\}}{\mathrm{E}_N\{N^2\}} = \frac{\sigma_n^2 - \sigma_{e_\mathrm{B}}^2}{\sigma_n^2} . \quad (16)$$

Plugging $\hat{n}_\mathrm{U} = 1/k_N \cdot \hat{n}_\mathrm{B}$ into (14) gives the unbiased estimate[3] for $x$

$$x_\mathrm{U} = z - \frac{1}{k_N} \cdot \hat{n}_\mathrm{B} = (1 - C_N) \cdot x_\mathrm{B} + C_N \cdot z , \quad (17)$$

with $C_N \stackrel{\text{def}}{=} \sigma_{e_\mathrm{B}}^2/(\sigma_{e_\mathrm{B}}^2 - \sigma_n^2)$. Due to construction, in this case the estimation error $E_\mathrm{U}$ is not orthogonal to $X$, but to $N$. Noteworthy, in contrast to the signal-based estimation where the unbiased estimate was a scaled version of $x_\mathrm{B}$, it depends on $x_\mathrm{B}$ as well as on $z$ if noise-based unbiasing is applied.

The unbiased error variance (w.r.t. $X$, $e_\mathrm{U} = x_\mathrm{U} - x$) can be calculated by

$$\varsigma_{e_\mathrm{U}}^2 = \mathrm{E}_X\{E_\mathrm{U}^2\} = (1 - C_N^2) \cdot \varsigma_{e_\mathrm{B}}^2 + C_N^2 \cdot \sigma_n^2 . \quad (18)$$

## 2.3. Discussion

Examples of the characteristic curves (top) of the biased and unbiased soft values, as well as the corresponding error variances (bottom), are given in Fig. 3 [12]. While the character-istic curves of the *biased* (blue) and the signal-based unbiased (green) estimates are strictly monotonically increasing, the noise-based *unbiased* curves (red) do not fulfill this property. As for all MMSE solutions, the unbiased estimates converge to the biased values if $\sigma_n^2$ tends to zero, since the deviation of the (biased) estimates from the correct value is negligible in this case.

---
[3]In [15], the (scaled) unbiased estimator is denoted as divergence-free.

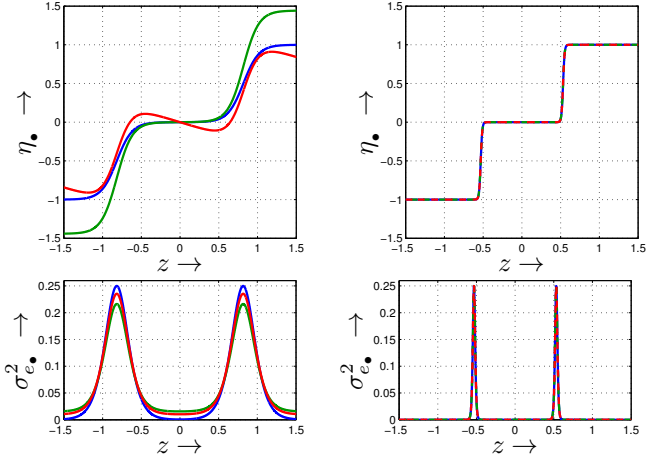

**Fig. 3**. Characteristic curves of biased (blue) and unbiased (green: signal-based, red: noise-based) soft values for $\sigma_n^2 = 0.1$ (left) and $\sigma_n^2 = 0.01$ (right). Soft values (top), error variance (bottom).

The variables present in the estimation process are visualized in Fig. 4, where the squared length of a vector corresponds to the variance of the corresponding variable [8]. The given variables $X$ and $N$ are orthogonal, and $Z = X + N$. In case of a *linear* estimator, $X_\mathrm{B}$ and $X_\mathrm{U}$ are scaled versions of $Z$, and $E_\mathrm{B} \perp Z$, cf. Fig. 4, left part. In case of signal-based unbiasing (red), $E_\mathrm{U}$ has to be orthogonal to $Z$, and hence $X_\mathrm{U} = Z$. For noise-based unbiasing (blue), $E_\mathrm{U}$ has to be orthogonal to $N$, and hence $X_\mathrm{U} = 0$.

The nonlinearity of soft feedback calculation introduces an additional degree of freedom, i.e., a three-dimensional signal space is needed to represent the variables. In particular, $X_\mathrm{B}$ will point out of the $X$-$N$-plane; this additional dimension, orthogonal to $X$ and $N$, is denoted as <u>v</u>ertical dimension $V$ in the following. The visualization in the right part of Fig. 4 is a two-dimensional projection of the three-dimensional graph. Due to the orthogonality conditions, $X_\mathrm{U}$ is an elongation of $X_\mathrm{B}$ on the $N$-$V$-plane for signal-based unbiasing; for noise-based unbiasing, $X_\mathrm{U}$ is determined by the intersection of an elongation of $\tilde{N}_\mathrm{B}$ with the $X$-$V$-plane.

Hence, while the unbiased error $E_\mathrm{U}$ remains in the $N$-$V$-plane in case of signal-based unbiasing, if noise-based unbiasing is applied it lays in the $X$-$V$-plane, i.e., $E_\mathrm{U}$ is orthogonal to the noise. If applied in an algorithm where linear estimation and soft-feedback calculation are iterated, in case of signal-based unbiasing the error is constantly in the $N$-$V$-plane. For noise-based unbiasing the error alternates between the $N$-$V$-

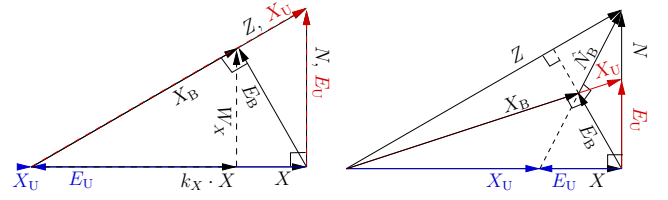

**Fig. 4**. Visualization of the variables present in the estimation process [12]. Linear estimation (left), nonlinear estimation (right). Red: Signal-based unbiasing. Blue: Noise-based unbiasing.

and $X$-$V$-plane every other iteration, which is beneficial for convergence and accuracy.

### 2.4. Connection to Average Variances

Often, the soft values of a *vector* $z$ have to be calculated. In this case, the unbiasing equations ((11) and (13) in the signal-based case, (17) and (18) for noise-based unbiasing) are applied for each element individually, leading to individual estimates and individual error variances.

However, in some situations, instead of *individual* variances, an *average* variance (averaged over the entire vector) should characterize the reliability. To this end, all elements of $z$ are assumed to have the same *average* noise variance $\sigma_n^2$. Then, $\varsigma_{e_\mathrm{B}}^2$ of the individual elements (Eq. (4)) has to be replaced by $\bar{\sigma}_{e_\mathrm{B}}^2$ (in practice, the expectation according to (5) is replaced by $\bar{\sigma}_{e_\mathrm{B}}^2$ that is averaging $\varsigma_{e_\mathrm{B}}^2$ over the vector elements), leading to the unbiasing equations (for the $l^\mathrm{th}$ element of the vector)

$$x_{\mathrm{U},l} = \sigma_{e_\mathrm{U}}^2 \cdot \frac{x_{\mathrm{B},l}}{\bar{\sigma}_{e_\mathrm{B}}^2}, \qquad \sigma_{e_\mathrm{U}}^2 = \left(\frac{1}{\bar{\sigma}_{e_\mathrm{B}}^2} - \frac{1}{\sigma_x^2}\right)^{-1},$$

for signal-based unbiasing, and

$$x_{\mathrm{U},l} = \sigma_{e_\mathrm{U}}^2 \cdot \left(\frac{x_{\mathrm{B},l}}{\bar{\sigma}_{e_\mathrm{B}}^2} - \frac{z_l}{\bar{\sigma}_n^2}\right), \qquad \sigma_{e_\mathrm{U}}^2 = \left(\frac{1}{\bar{\sigma}_{e_\mathrm{B}}^2} - \frac{1}{\bar{\sigma}_n^2}\right)^{-1},$$

in the noise-based case. Note that the latter equations equal the ones used in [13, 14, 15] without any justification, in particular not the above given interpretation.

## 3. APPLICATION TO COMPRESSED SENSING

In Compressed Sensing, a sparse vector $x$ has to be estimated from an underdetermined system of linear equations, which is given by [16]

$$y = Ax + w, \qquad (19)$$

where the received vector $y \in \mathbb{R}^K$ depends on the measurement (channel) matrix $A \in \mathbb{R}^{K \times L}, L > K$, and on the sparse vector $x \in \mathcal{C}^L$ (with sparsity $s$), where $\mathcal{C} \subseteq \mathbb{R}$. Furthermore, the measurements are corrupted by i.i.d. zero-mean Gaussian noise $w$ with variance $\sigma_w^2$ per component, which is independent of $x$. In *discrete* Compressed Sensing, the elements of

$\boldsymbol{x}$ are furthermore assumed to be drawn from a finite set, i.e., $\mathcal{C} = \{0, c_1, \ldots, c_{|\mathcal{C}|-1}\}$.

Due to the sparsity constraint and the discrete alphabet, the problem of estimating $\boldsymbol{x}$ based on (19) is non-convex. Different algorithms for the approximate solution of the problem are available in the literature; for a detailed discussion thereon, cf., e.g., [13, 14].

### 3.1. Algorithm

In [13], an iterative algorithm denoted as IMS has been proposed which splits the estimation problem into two parts, alternatingly estimating $\boldsymbol{x}$ w.r.t. the Gaussian noise (Linear MMSE estimation, index $\cdot_{\text{L}}$), and w.r.t. $s$ and $\mathcal{C}$ (Nonlinear MMSE estimation (soft feedback), index $\cdot_{\text{N}}$)). The pseudocode of this algorithm is given in Alg. 1.

While the MMSE estimate in the first step is unbiased, the soft values in the second step are not unbiased in the original algorithm (i.e., biased, Line 5B). Thus, we apply the equations for individual signal- or noise-based unbiasing derived in this paper and denote the new algorithm, including the unbiasing, as xuIMS and nuIMS, respectively (cf. Alg. 1, Line 5U).

Note that the TMS algorithm [14] (strongly related to OAMP or VAMP [15, 17]) is similar to nuIMS, however using *average* variances instead of *individual* ones; thus, it does not benefit from the information about the reliability of the particular elements as does uIMS.

**Alg. 1** $\hat{\boldsymbol{x}} = $ recover $(\boldsymbol{y}, \boldsymbol{A}, \sigma_w^2, s, \mathcal{C})$
Variants: B: IMS, U: $(\cdot)$uIMS

1 : $\boldsymbol{x}_{\text{L,U}} = \boldsymbol{0}, \sigma_{e_{\text{N,U}},l}^2 = s/L \;\forall l$
2 : **while** stopping criterion not met {
3 : $(\boldsymbol{x}_{\text{L,U}}, \boldsymbol{\sigma}_{e_{\text{L,U}}}^2) = $ unbiased linear MMSE estimate $(\boldsymbol{A}, \boldsymbol{y}, \boldsymbol{x}_{\text{N,U}}, \boldsymbol{\sigma}_{e_{\text{N,U}}}^2)$
4 : $(x_{\text{N,B},l}, \varsigma_{e_{\text{N,B},l}}^2) = $ biased soft feedback $(x_{\text{L,U},l}, \sigma_{e_{\text{L,U},l}}^2, s, \mathcal{C})$
5B: $(x_{\text{N,U},l}, \sigma_{e_{\text{N,U},l}}^2) = (x_{\text{N,B},l}, \varsigma_{e_{\text{N,B},l}}^2)$
5U: $(x_{\text{N,U},l}, \sigma_{e_{\text{N,U},l}}^2) = $ unbias$(x_{\text{N,B},l}, \varsigma_{e_{\text{N,B},l}}^2)$
6 : }
7 : $\hat{\boldsymbol{x}} = \mathcal{Q}_{\mathcal{C}}(\boldsymbol{x}_{\text{N,B}})$

### 3.2. Simulation Results

The performance of IMS with unbiasing is shown in Fig. 5 for $L = 258$, $K = 129$, $s = 15$, $\mathcal{C} = \{-1, 0, +1\}$. The measurement matrix is a random Gaussian matrix, with the columns normalized to unit norm. In the upper part, the symbol error rate (SER) over the noise variance is shown. In order to ensure convergence, all algorithms perform 50 iterations. Besides IMS and $(\cdot)$uIMS, also the results for noise-based unbiasing with *average* variances (TMS [14]) and for the BAMP algorithm [18] are shown.

While IMS without unbiasing (green) performs even worse than TMS (blue) which tracks only average instead of individual variances, the performance can be improved if individual signal-based unbiasing (xuIMS, red dashed) is applied. Individual noise-based unbiasing (nuIMS, red solid), however, clearly outperforms the other algorithms by $0.7$ dB and $0.5$ dB, respectively. Hence, the disadvantage that the unbiased error is not orthogonal to the signal, is overcompensated by the orthogonality of the noise/input error and the estimation error (cf. Sec. 2.3). Furthermore, the BAMP algorithm (black), which is state of the art in Compressed Sensing, is also clearly outperformed.

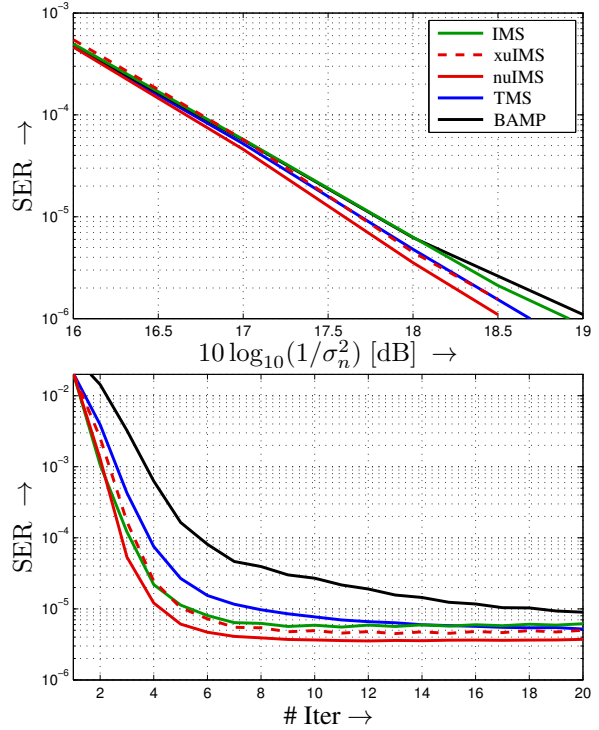

**Fig. 5**. SER of the proposed algorithm with unbiasing over the noise level $1/\sigma_n^2$ in dB (top), and over the iterations for $10 \log_{10}(1/\sigma_n^2) = 18$ dB (bottom). $L = 258$, $K = 129$, $s = 15$, $\mathcal{C} = \{-1, +1\}$. Gaussian matrix, $\|\boldsymbol{A}(:,l)\|^2 = 1 \,\forall l$.

In the lower part if Fig. 5, the SER over the number of iterations is shown for $10 \log_{10}(1/\sigma_n^2) = 18$ dB. Noteworthy, IMS and both variants of uIMS converge significantly faster than TMS, i.e., the tracking of individual instead of average variances does not only improve the performance, but accelerates also the convergence. Furthermore, BAMP performs again worst.

## 4. CONCLUSION

In this paper, we have discussed the unbiasing for soft feedback, treating the estimation problem from the signal as well as from the noise perspective. Both approaches have been compared, and the connection to solutions with average variances, as they are widely used in the literature, has been pointed out. Furthermore, both derived unbiasing variants have been employed in an iterative algorithm, and the gains due to the unbiasing have been investigated by numerical results.